\documentclass[aps,prb,preprint,12pt]{revtex4-1}

\usepackage[applemac]{inputenc}
\usepackage[T1]{fontenc}
\usepackage{graphicx}
\usepackage{amsmath,amssymb}
\usepackage{cases}

\begin{document}

\title{Heat capacity study of the magnetic phases in a Nd$_5$Ge$_3$ single crystal}

\author{D. Villuendas}
\email[Corresponding author: ]{diego@ubxlab.com}
\affiliation{Facultat de F\'isica, Universitat de Barcelona, Mart\'i i Franqu\`es 1, 08028 Barcelona, Spain}

\author{J. M. Hern\`andez}
\affiliation{Facultat de F\'isica, Universitat de Barcelona, Mart\'i i Franqu\`es 1, 08028 Barcelona, Spain}

\author{T. Tsutaoka}
\affiliation{Graduate School of Education, Hiroshima University, Higashi-Hiroshima, Hiroshima 739-5824, Japan}

\date{\today}

\begin{abstract}
The different magnetic phases of the intermetallic compound Nd$_5$Ge$_3$ are studied in terms of the specific heat, in a broad range of temperatures (350~mK--140~K) and magnetic fields (up to 40 kOe). 
The expected $T^{3}$ and  $T^{3/2}$ terms are not found in the antiferromagnetic (AFM) and ferromagnetic (FM) phases respectively, but a gapped $T^2$ contribution that originates from a mixture of AFM and FM interactions in different dimensionalities under a large magnetocrystalline anisotropy, is present in both. An almost identical Schottky anomaly, that arises from the hyperfine splitting of the nuclear levels of the Nd$^{3+}$ ions, is observed in both phases, which leads us to state that the magnetic-field induced transition AFM$\to$FM that the system experiments below $26~\text{K}$ consists in the flip of the magnetic moments of the Nd ions, conserving the average local moment.\end{abstract}


\maketitle

\section{Introduction}
In the last years the binary intermetallic compound Nd$_5$Ge$_3$ has been the object of different studies. The interest in this system comes from the fact that many of its physical properties present abrupt changes when the transition between its two magnetic phases occurs. The first investigations of the magnetic properties showed that this material orders ferrimagnetically in zero applied magnetic field at the N\'eel temperature, $T_{N}\approx50~\text{K}$, and possesses a remanent moment at $4.2~\text{K}$~\cite{Buschow1967}. Later on,  neutron diffraction experiments suggested that when the system is cooled in zero applied magnetic field an antiferromagnetic (AFM) state is established below $T_{N}$, although the hysteresis loop typical of a hard magnetic material was found at $4.2~\text{K}$. This fact indicated a magnetic-field-induced phase transition to a ferromagnetic (FM) state~\cite{Schobinger-Papamantellos1985}. In recent years, the system has regained attention fostered by the exploration of the temperature dependence of the irreversibility of this transition below $T_{t}=26~\text{K}$~\cite{Tsutaoka2010a}. Some papers have been published on this compound showing rich phenomena in magnetostriction~\cite{Doerr2009}, electric resistance, specific heat, spontaneous magnetic phase transitions~\cite{Maji2010}, and more recently optical properties and electronic structure~\cite{Knyazev2014}.

The intermetallic alloy Nd$_5$Ge$_3$ belongs to the family R$_5$Ge$_3$, with R~=~rare earth, that crystallizes in the hexagonal Mn$_5$Si$_3$-type structure ($P6_{3}/mcm$, space group No.~193). The structure contains two formula units per unit cell, in which R atoms occupy the two non-equivalent crystallographic sites $4d$ and $6g$, while Ge atoms occupy the $6g$ site~\cite{Zeng2000}. Although it is still unclear, the most accepted magnetic structure, which has been derived from neutron diffraction experiments~\cite{Schobinger-Papamantellos1985,Nirmala2011,Vokhmyanin2014}, consists in a complex double sheet, with the magnetic moments of the Nd ions located in the $6g$ position oriented parallel to the $c$-axis, and the moments of the Nd ions in the $4d$ position oriented along the $c$-axis with a deviation angle of 31$^{\circ}$ and a propagation vector $\mathbf{k}=(0.25\,0\,0)$; the $z$ component changes sign every two successive (110) planes.\\

The temperature variation of the specific heat in zero field was measured in Ref.~\onlinecite{Tsutaoka2010a} and ~\onlinecite{Maji2011} to examine the magnetic phase transitions. Both works observed a hump around $50~\text{K}$ but no anomaly around $26~\text{K}$, which was attributed to the existence of a spin-glass state, because of the occurrence of similar features in the $C_{p}(T)$ curve in other well-known spin-glass systems. In this paper we will conduct a detailed study of the specific heat, as a function of the temperature and the magnetic field, seeking a better understanding of the properties of the different magnetic phases.

\section{Methods}
Polycrystalline ingots were prepared by arc-melting the constituting 99.9\%-pure Nd and 99.999\%-pure Ge elements under high-purity argon atmosphere. The compounds were found to be single-phase by powder X-ray diffraction. Single crystals were grown by the Czochralski method from single-phase polycrystalline samples using a tri-arc furnace. It should be noted that it is difficult to grow a large single crystal of Nd$_5$Ge$_3$ because the grown crystal ingots tend to have small single-crystalline grains. The sample was cut from one ingot into a rectangular shape ($1\times 1.5\times 2~\text{mm}^3$) and annealed at 300$^{\circ}$C for 24~h in an evacuated quartz tube. The crystal orientation was determined by the back-reflection Laue method. Measurements of the specific heat in the temperature range from 350~mK to 300~K and magnetic fields up to 40~kOe were made using the heat pulse-relaxation method with the heat capacity option of the PPMS$^{\circledR}$ system, produced by Quantum Design$^{\circledR}$.

\section{Results and discussion}

\begin{figure}
 \includegraphics{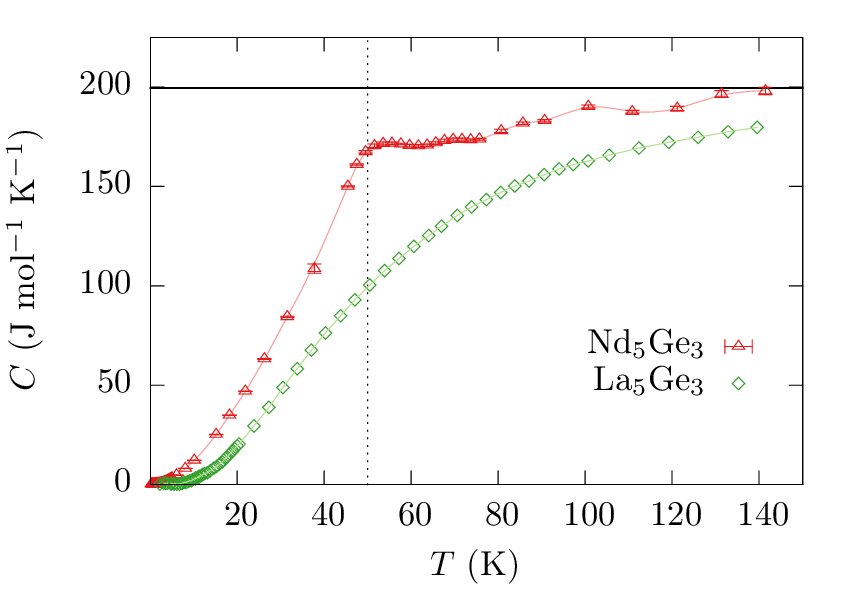}%
 \caption{\label{fig:cT}(Color online) Zero-field temperature dependence of the specific heat of a single crystal of Nd$_5$Ge$_3$ (triangles) and a polycrystal of La$_5$Ge$_3$ (diamonds; extracted from Joshi \emph{et al.}~\cite{Joshi2009}). The lines joining the data points are guides to the eye. The vertical dotted line indicates the temperature (50~K) at which a hump is observed for Nd$_5$Ge$_3$. The horizontal solid line indicates the Dulong-Petit limit of the specific heat.}
 \end{figure}

Figure~\ref{fig:cT} shows the specific heat, $C$, of our Nd$_5$Ge$_3$ single crystal (triangles) and a polycrystaline sample of the  nonmagnetic isostructural compound La$_5$Ge$_3$ (diamonds; extracted form Joshi \emph{et al.}~\cite{Joshi2009}) as a function of temperature. The specific heat of the latter will be used as a blank and follows the expected monotonic behavior from the electronic and phononic contributions~\cite{Kittel2005}, while the specific heat of Nd$_5$Ge$_3$ presents a hump around $T_{N}\simeq50~\text{K}$ and thenceforth tends progressively to the Dulong-Petit limit.

\begin{figure}
 \includegraphics{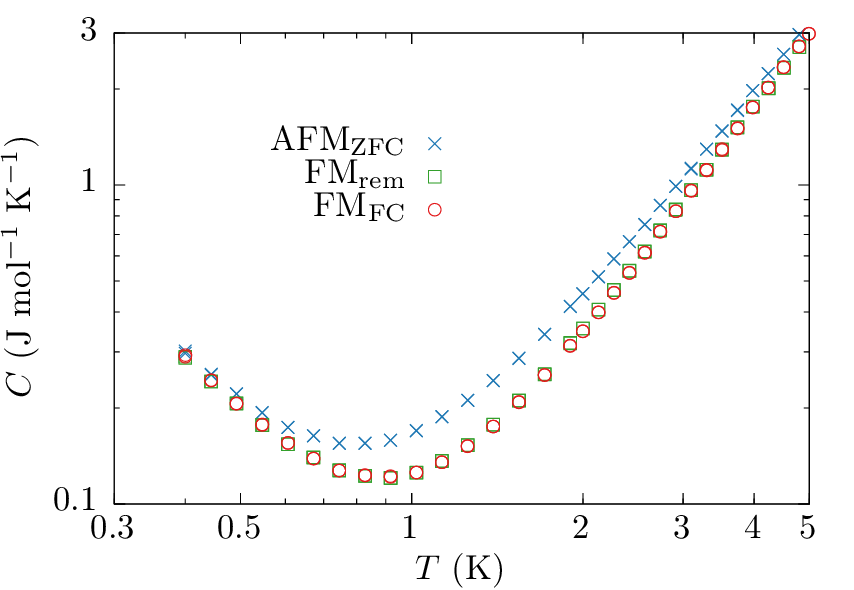}%
 \caption{\label{fig:afremfc}(Color online) Log-log plot of the low-temperature dependence of the specific heat of the three magnetic states: the AFM (crosses), the FM$_{\text{rem}}$ (squares), and the FM$_{\text{FC}}$ (circles).}
 \end{figure}

The dependence of the specific heat of Nd$_5$Ge$_3$ with the temperature was investigated as follows. 
Following a zero-field cooling process (ZFC) the specific heat was measured from $140~\text{K}$ down to $350~\text{mK}$. During this process we observed the expected hump associated to the paramagnetic--AFM transition  at 50~K. The next step was to measure the system in the FM state. It is known that cooling the Nd$_5$Ge$_3$ compound below $26~\text{K}$ with an applied magnetic field larger than $5~\text{kOe}$ leaves the system in the saturated FM state~\cite{Tsutaoka2010a}. To ensure this fact we measured the specific heat following a field cooling process (FC) in an applied field of $15~\text{kOe}$, large enough for our purpose. Nonetheless our goal was to compare the magnetic contribution of the different magnetic phases to the specific heat, and the applied magnetic field could play an undesired role. Therefore, after measuring the FC process we set the magnetic field to zero and measured the specific heat of the ferromagnetic remanent state (FM$_{\text{rem}}$) as we increased the temperature. From magnetization experiments it is known that well below $26~\text{K}$ the FM$_{\text{rem}}$ and the FM$_{\text{FC}}$ are magnetically equivalent~\cite{Tsutaoka2010a}. We observed that this equivalence is also present in terms of specific heat, as it is shown in Fig.~\ref{fig:afremfc}. In this figure the three data sets are plotted together showing the AFM curve and how the FM$_{\text{rem}}$ and FM$_{\text{FC}}$ curves superimpose. Consequently, from now on we will rename the FM$_{\text{rem}}$ state as FM state in this temperature region. 

The specific heat can be assumed to be made up of four independent contributions, 
\begin{equation}
C(T)=C_{\text{el}}(T)+C_{\text{lat}}(T)+C_{\text{hyp}}(T)+C_{\text{mag}}(T).
\label{eq:4contributions}
\end{equation}

The contribution from phonons, $C_{\text{lat}}$, can be subtracted using the specific heat of the nonmagnetic isostructural compound La$_5$Ge$_3$, taking into account the different molar masses of Nd and La via  the two-Debye function method~\cite{Bouvier1991,Hoffmann1956},
\begin{align}
\label{eq:twoDebye}
C_{\text{lat}}^{\text{Nd}_5\text{Ge}_3}(T)=C_{\text{lat}}^{\text{La}_5\text{Ge}_3}(r\,T),\\
\intertext{with}
r=\left(\frac{5M_{\text{La}}^{3/2}+3M_{\text{Ge}}^{3/2}}{5M_{\text{Nd}}^{3/2}+3M_{\text{Ge}}^{3/2}}\right)^{1/3}=0.98.
\end{align}
Therefore we get
\begin{equation}
C(T)-C_{\text{lat}}^{\text{La}_5\text{Ge}_3}(r\, T)=C_{\text{el}}(T)+C_{\text{hyp}}(T)+C_{\text{mag}}(T).
\label{eq:4contributions-La}
\end{equation}

\begin{figure}
 \includegraphics{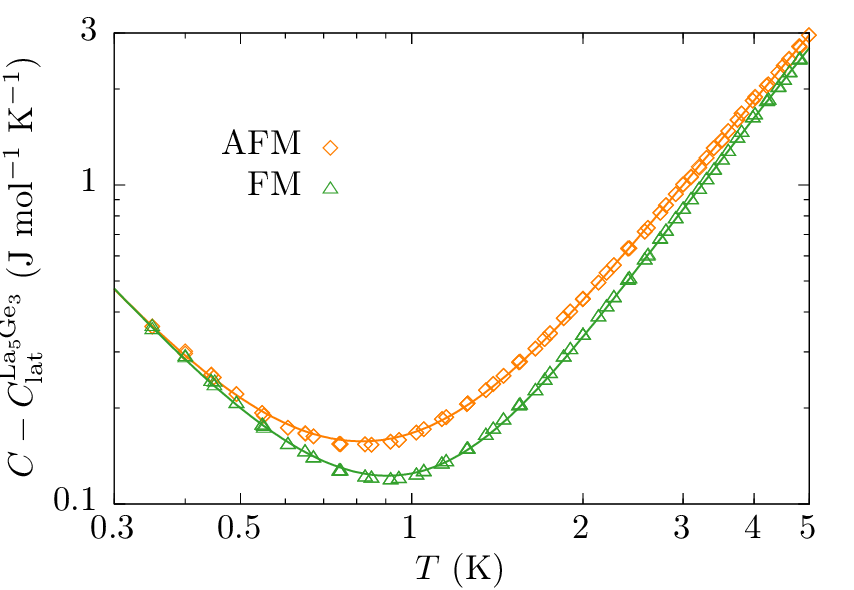}%
 \caption{\label{fig:fitaffrem}(Color online) Zero-field low-temperature specific heat for both the AFM (diamonds) and FM (triangles) states, plotted in a log-log scale to remark the differences between both states. The lattice contribution to the specific heat has been subtracted to the experimental data. The lines are the best fits of Eq.~\eqref{eq:4contributions-La} to the data.}
 \end{figure}

To determine the $C_{\text{mag}}(T)$ contribution for each magnetic phase, we can attempt to model the experimental data taking into account the different terms: $C_{\text{el}}(T)=\gamma T$ from free charge carriers,  $C_{\text{hyp}}(T)=A T^{-2}$ from the high-temperature limit of the Schottky anomaly due to the hyperfine splitting of the nuclear levels of the Nd$^{3+}$ ions, and $C_{\text{mag}}(T)$ from spin waves. 
The approach to study the last term was to consider the more general dispersion relations for the long-wavelength spin interactions. We examined the cases of AFM, FM, and type-$A$ AFM (ferromagnetic layers antiferromagnetically coupled) states. The last is one of the proposed magnetic structures to occur below $T_N$ at zero applied magnetic field~\cite{Schobinger-Papamantellos1985}. Because of the large magnetic anisotropy and the magnetoelastic effects present in this system~\cite{Tsutaoka2010a,Doerr2009}, we also took into account the possibility of the presence of a gap in the dispersion relation, $\Delta$. In the low-temperature limit, the specific heat from each dispersion relation is found to be
\begin{subnumcases}{C_{\text{mag}}(T)=B e^{-\frac{\Delta}{T}}}
T(12T^2+6T\Delta+\Delta^2)\label{eq:CAFM}\\
T^{-\frac{1}{2}}(15 T^2+12T\Delta+4\Delta^2)\label{eq:CFM}\\
(6T^2+4T\Delta+\Delta^2)\label{eq:CFM2AFM1}.
\end{subnumcases}
Equations~\eqref{eq:CAFM}, \eqref{eq:CFM} and \eqref{eq:CFM2AFM1} correspond, respectively, to the low-temperature magnetic contribution to the specific heat of the AFM, the FM, and the type-$A$ AFM states. The usual expressions ($T^3$, $T^{3/2}$ and $T^2$, respectively) are recovered when $\Delta$ is set to zero. We fitted the experimental data to the expressions with and without gap and found that, neither in the AFM nor in the FM phase, the ``pure'' AFM/FM contributions gave physically reasonable values for the parameters. On the contrary, in both phases the gapped type-$A$ AFM contribution [Eq.~\eqref{eq:CFM2AFM1}] was found to fit precisely. Figure~\ref{fig:fitaffrem} shows the specific heat for the AFM and FM states together with the best fits of Eq.~\eqref{eq:4contributions-La} to the data. Table~\ref{tab:fitaffrem} lists the coefficients of all contributions. 

 \begin{table}
\caption{\label{tab:fitaffrem}Results of the fitting of Eq.~\eqref{eq:4contributions-La} to the specific heat data, for both the AFM and FM states. The units are mJ$/$(mol K$^{m+1})$ where $m$ is the power of $T$ corresponding to each coefficient, $m=1$ for $\gamma$, $m=-2$ for $A$, $m=2$ for $B$. The number in parentheses is the statistical uncertainty in the last digit from the least-squares fitting procedure.}
\begin{ruledtabular}
\begin{tabular}{*{5}{c}}
state & $\gamma$ & $A$ & $B$ & $\Delta$~(K) \\
 \hline
AFM & 115(2) & 39.3(4) & 22.2(1) & 4.34(8)  \\
FM & 75(1) & 40.6(3) & 23.1(1) & 4.75(5)  \\
\end{tabular}
\end{ruledtabular}
\end{table}
 
The mixture of interactions and dimensionalities resulting in a $T^2$ contribution to $C_\text{mag}$ has been also considered in magnetic structures where FM droplets are found in an AFM phase~\cite{He2009}. In our system, nevertheless, the magnetization measurements indicate a saturated FM state, without  evidence of AFM interactions. Despite we do not have a clear explanation to the presence of this term in the FM phase, the values for the parameters obtained fitting other contributions [Eq.~\eqref{eq:CAFM} and \eqref{eq:CFM}] do not have any physical meaning. The larger gap obtained in the FM state corresponds with the larger internal magnetic field in this phase, as it will be shown below. The values of $\gamma$ for both phases are reasonable within the electronic contributions of rare earth intermetallics R$_5$Ge$_3$~\cite{Gorbachuk2010} and do not need extra considerations. The reduction of the value in the FM phase with respect to the AFM can be attributed to a decrease in the density of states at the Fermi level, that would probably favour one of the electronic spin projections. 

\begin{figure}
 \includegraphics{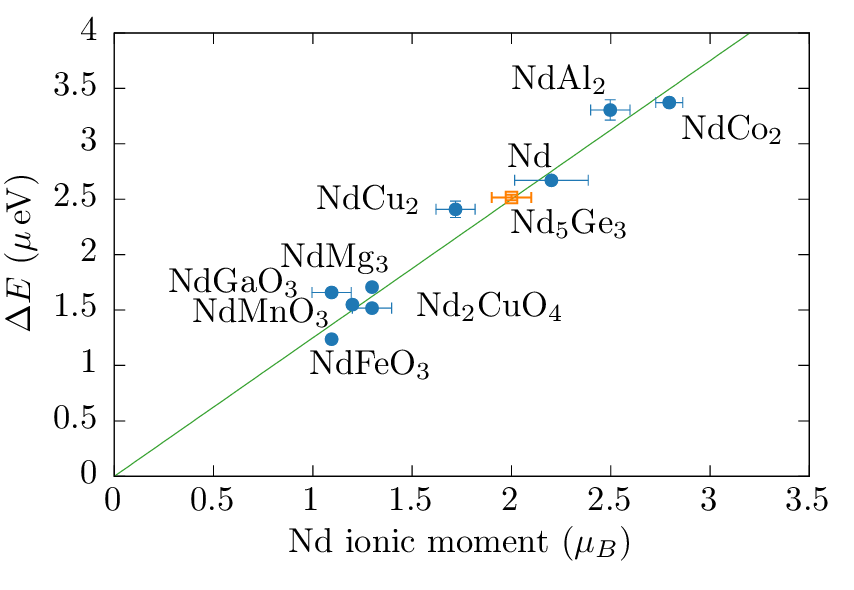}%
 \caption{\label{fig:Evsmu}(Color online) Energy splitting of inelastic neutron scattering signals in several Nd-based compounds  (circles) as a function of the corresponding saturated ionic magnetic moment of Nd at low temperatures (adapted from Chatterji \emph{et al.}~\cite{Chatterji2009}). The data point obtained in this work for Nd$_5$Ge$_3$ (square) is also shown.}
 \end{figure}

The hyperfine contribution can not be omitted to fit completely the measured specific heat. The Schottky anomaly consists in a peak originated from the (de)population of discrete energy levels. In this case, these correspond to the hyperfine split nuclear levels of the Nd$^{3+}$ ions. The Schottky anomaly can be approximated to $A/T^{2}$ in its high-temperature limit, where $A$ is related to the internal hyperfine magnetic field by the expression~\cite{He2009}
\begin{equation}
\label{eq:schottky}
A=5\frac{N_{A}k_{B}}{3}\left(\frac{I+1}{I}\right)\left(\frac{\mu_{I}H_{\text{hyp}}}{k_{B}}\right)^{2}.
\end{equation} 
Here $I$ is the nuclear spin, $\mu_{I}$ is the nuclear magnetic moment, $H_{\text{hyp}}$ is the internal magnetic field at the Nd site, and the factor 5 is the number of moles of Nd per mole of Nd$_5$Ge$_3$. Only two isotopes of Nd have nuclear spin different from zero ($I=7/2$),  $^{143}$Nd and $^{145}$Nd with the natural abundances of 12.18\% and 8.29\%, whose nuclear magnetic moments are $\mu_I=-1.208\mu_{N}$ and $\mu_I=-0.744\mu_{N}$ respectively~\cite{Harris1996}. The hyperfine field values obtained are $\mu_{0}H_{\text{hyp}}(\text{AFM})=272(2)~\text{T}$ and $\mu_{0}H_{\text{hyp}}(\text{FM})=276(1)~\text{T}$. The energy splitting ($\Delta E=\mu_{I}H_{\text{hyp}}/I$) found is $\sim 2.5~\mu\text{eV}$ for both phases.  We can compare this value with the splitting of other Nd compounds studied with neutron scattering. Figure~\ref{fig:Evsmu} plots the hyperfine energy splitting versus the saturated ionic magnetic moment of Nd for several Nd-based compounds~\cite{Chatterji2009}, along with the data point obtained in this work. The value used for the magnetic moment of the Nd ion corresponds to the one observed in the saturated FM state~\cite{Tsutaoka2010a}, $\mu_{\text{Nd}}\approx 2\mu_{B}$. It is remarkable that we have obtained approximately the same splitting for both magnetic phases, $\Delta E(\text{AFM})=2.50(2)~\mu\text{eV}$ and $\Delta E(\text{FM})=2.53(1)~\mu\text{eV}$, which means that the average local magnetic moment per Nd ion is roughly the same in the two phases and corresponds to the value in the FM saturated state. Therefore, we may assert that the magnetic-field-induced AFM$\to$FM transition simply flips the magnetic moments and preserves the value of $\mu_\text{Nd}$.

We will now proceed to investigate the temperature dependence of the magnetic contribution to the specific heat. Subtracting to the total measured specific heat the analytical functions of the electronic contribution, the Schottky anomaly and the phononic contribution from the corrected  La$_5$Ge$_3$ data we obtain $C_{\text{mag}}(T)$. 
\begin{figure}
 \includegraphics{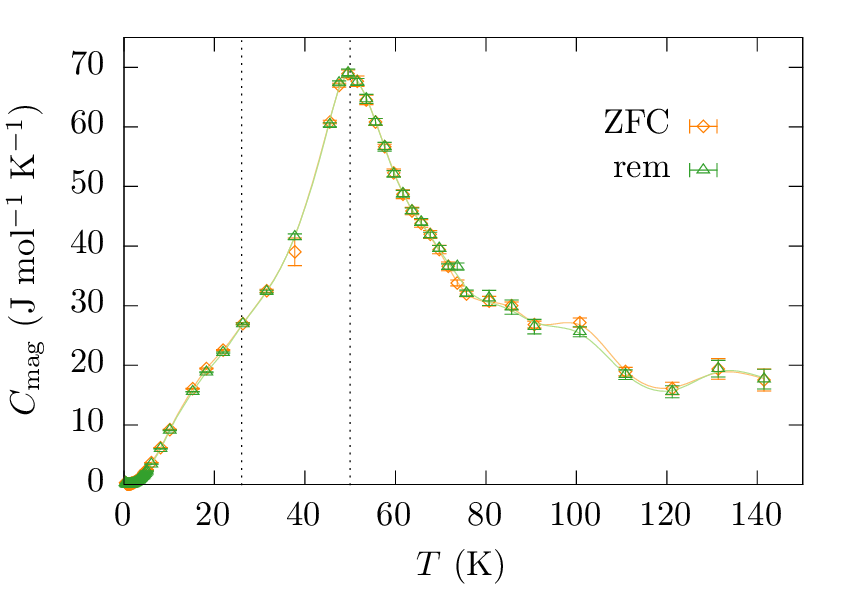}%
 \caption{\label{fig:cmag}(Color online) Temperature dependence of the magnetic contribution to the specific heat obtained following the ZFC (diamonds) and rem (triangles) processes. A maximum at 50~K and an inflection point around 26~K are highlighted with vertical dotted lines. The lines joining the data points are guides to the eye.}
 \end{figure}
Fig.~\ref{fig:cmag} shows the zero-field-cooled (ZFC) and remanent (rem) curves, where the latter was acquired as the sample was heated in zero applied magnetic field after it had been driven to the FM saturated state. The peak at 50~K indicates the temperature at which the AFM ordering takes place ($T_{N}$). 
The inflection point observed around 26~K is in contradiction with the previously reported absence of any anomaly around this temperature, which was related to the possibility of the system to be in a spin glass state~\cite{Tsutaoka2010a,Maji2011}.  

From $C_{\text{mag}}(T)$ we can compute the magnetic entropy as
\begin{equation}
\label{eq:entropy}
S_{\text{mag}}(T)=\int_{0}^{T}\frac{C_{\text{mag}}(T')}{T'} \text{d}T',
\end{equation}
  \begin{figure}
 \includegraphics{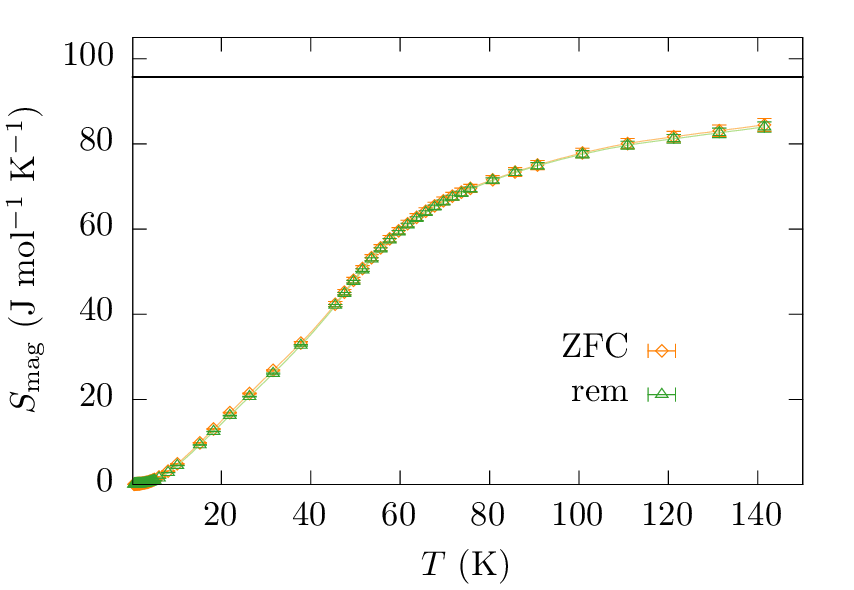}%
 \caption{\label{fig:entropy}(Color online) Temperature dependence of the magnetic entropy obtained for the ZFC (diamonds) and rem (triangles) processes. The horizontal solid line represents the value of $5R\ln(2J+1)$ for $J=9/2$. The lines joining the data points are guides to the eye.}
 \end{figure}
where it is assumed that the magnetic entropies of AFM and FM materials at zero temperature are zero. Fig.~\ref{fig:entropy} shows how $S_\text{mag}$ attains the value of $R\ln(2J+1)$ expected for a paramagnet~\cite{Blundell2001} as the temperature grows above $T_{N}$. In our case the entropy tends to $5R\ln10$, corresponding to 5 Nd$^{3+}$ free ions with $J=9/2$. The actual value at which the obtained entropy tends is moderately smaller because the zero-field splitting due to the anisotropy could play a significant role even in the paramagnetic state.

Finally, the dependence of the specific heat with the applied magnetic field at fixed temperature was studied to explore the magnetic-field-induced AFM$\to$FM transition . The system was prepared following a ZFC process from $T\gg T_{\text{N}}$ down to 1.2~K, where the relative difference between the specific heat of the two phases has a maximum, and then the specific heat was measured varying the field, from 0 to $40~\text{kOe}$ and back to $0~\text{Oe}$.  
\begin{figure}
 \includegraphics{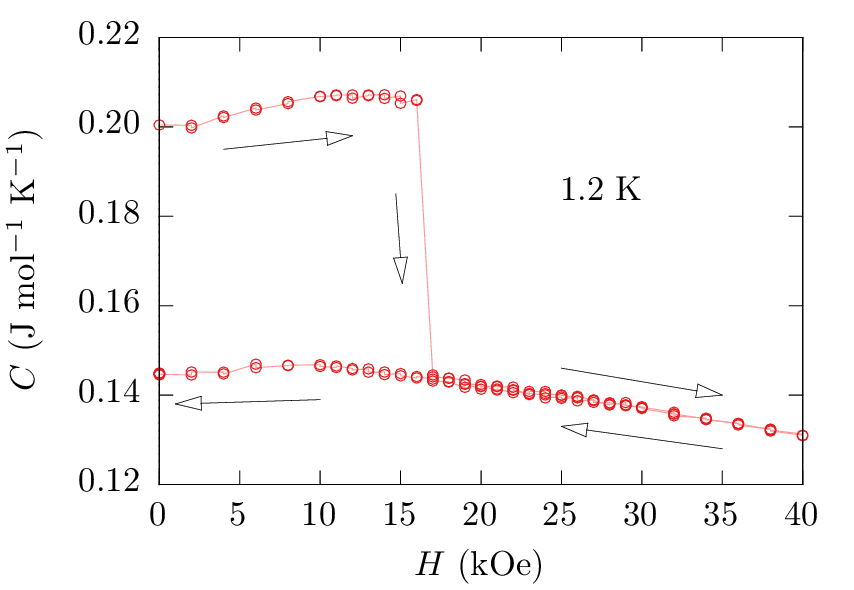}%
 \caption{\label{fig:cH}(Color online) Magnetic field dependence of the specific heat at 1.2~K, starting with the system in the AFM state ($C\approx 0.2~\text{J~mol}^{-1}~\text{K}^{-1}$) and ending with the system in the FM state ($C\approx 0.14~\text{J~mol}^{-1}~\text{K}^{-1}$). Two independent runs are plotted. The lines joining the data points are guides to the eye.}
 \end{figure}
Fig.~\ref{fig:cH} shows how a large, abrupt, and irreversible change in $C(H)$, associated with the AFM--FM transition, takes place between 16 and 17~kOe as the magnetic field ncreases. Two independent runs are plotted in the figure showing the reproducibility of this transition. From the magnetic field-temperature phase diagram obtained from magnetization measurements in Ref.~\onlinecite{Tsutaoka2010a} one expects this transition to happen at much higher fields $(H\gtrsim 35~\text{kOe})$. One explanation for this reduction of the field at which the spontaneous transition occurs is to consider it of thermally assisted origin. The large difference in the thermal bath properties between the two experimental setups (MPMS$^{\circledR}$ for magnetic measurements, versus PPMS$^{\circledR}$ for specific heat measurements) can strongly affect how thermally-assisted abrupt transitions develop~\cite{Webster2007,Macia2009}. We also see in the figure that the decreasing dependence of the specific heat with the increasing magnetic field, above $H\approx10~\text{kOe}$, is consistent with the behaviour of a $C_\text{mag}$ term with a gap in the dispersion relation of the spin waves proportional to the applied magnetic field, $\Delta\propto H$. Nevertheless, a change in the sign of the slope is observed in both states around $10~\text{kOe}$, for which we do not have an explanation. A more detailed study considering also magnetoelastic effects could give a better description of the exact behaviour of the specific heat as a function of the applied magnetic field.

\section{Conclusions}
In summary, we have performed measurements of the specific heat in the two magnetic phases of the system Nd$_5$Ge$_3$. From the low-temperature data we have modeled the different contributions to the specific heat. A magnetic $T^{2}$ contribution is found in both the ferromagnetic (FM) and antiferromagnetic (AFM) phases. This term can be understood as a mixture of FM and AFM interactions in different dimensionalities. In the case of the AFM phase this $T^2$ term can be attributed to a type-$A$ AFM, while in the case of the FM phase can be interpreted as a remanence of AFM interactions. The large magnetocrystalline anisotropy of Nd$_5$Ge$_3$ is evidenced by a gapped spin-wave spectrum in both phases.

The average magnetic moment at low temperature in the two magnetic phases has been obtained by means of the specific heat contribution of the hyperfine splitting of the nuclear moment of the Nd$^{3+}$ ions. The value of this magnitude is approximately $2\mu_{B}$ in both phases, which corresponds to the saturation value of the FM state at low temperature. Hence, we state that the magnetic-field-induced transition between both states corresponds to an irreversible spin-flip transition of the Nd ions.

Finally, from the magnetic field dependence we observe that the field at which the spontaneous transition takes place is remarkably smaller than the expected value from the magnetic field-temperature phase diagram. This is most likely due to the effect of being in an environment with a smaller thermal coupling (PPMS$^{\circledR}$ vs MPMS$^{\circledR}$), leading to a spontaneous ignition of a thermally assisted transition at smaller fields, probably by means of a magnetic deflagration process.

\begin{acknowledgments}
This work was financially supported by Spanish Government project MAT2011-23698. Authors would like to acknowledge the use of Servicio General de Apoyo a la Investigaci\'on-SAI, Universidad de Zaragoza. D. V. and J. M. H. also thank A. Garc\'ia-Santiago (UB) for useful discussions.
\end{acknowledgments}

\bibliography{/Users/diego/UBXLAB/Dropbox/Bibliography/Diego.bib}

\end{document}